\documentclass[prd,reprint,nofootinbib,amsmath,amssymb,aps,floatfix]{revtex4-1}
\usepackage{bm}% bold math
\usepackage{amssymb,amsmath,graphicx}
\usepackage[linktoc=page]{hyperref}
\usepackage{subfigure}
\usepackage{epsfig}
\usepackage{amsfonts}
\usepackage{color}
\usepackage[normalem]{ulem}

\begin{document}
%\normalem
\title{Information Content of Gravitational Radiation and the Vacuum}
\author{Raphael Bousso, Illan Halpern, and Jason Koeller}%
 \email{bousso@lbl.gov, illan@berkeley.edu, jkoeller@berkeley.edu}
\affiliation{ Center for Theoretical Physics and Department of Physics\\
University of California, Berkeley, CA 94720, USA 
}%
\affiliation{Lawrence Berkeley National Laboratory, Berkeley, CA 94720, USA}
\begin{abstract}

Known entropy bounds, and the Generalized Second Law, were recently shown to imply bounds on the information arriving at future null infinity. We complete this derivation by including the contribution from gravitons. We test the bounds in classical settings with gravity and no matter. In Minkowski space, the bounds vanish on any subregion of the future boundary, independently of coordinate choices. More generally, the bounds vanish in regions where no gravitational radiation arrives. In regions that do contain Bondi news, the bounds are compatible with the presence of information, including the information stored in gravitational memory. All of our results are consistent with the equivalence principle, which states that empty Riemann-flat spacetime regions contain no classical information. We also discuss the possibility that Minkowski space has an infinite vacuum degeneracy labeled by a choice of Bondi coordinates (a classical parameter, if physical). We argue that this degeneracy cannot have any observational consequences if the equivalence principle holds. Our bounds are consistent with this conclusion.

\end{abstract}
%\pacs{}
\maketitle
\tableofcontents

\section{Introduction}
\label{intro}

Entropy bounds control the information flow through any light-sheet~\cite{CEB1}, in terms of the area difference between two cuts $\sigma_1$, $\sigma_2$ of the light-sheet:
\begin{equation}
S\leq \frac{A[\sigma_1]-A[\sigma_2]}{4G\hbar}~.
\label{eq-genericbound}
\end{equation}
A light-sheet is a null hypersurface consisting of null geodesics orthogonal to $\sigma_1$ that are nowhere expanding. A cut is a spatial cross-section of the light-sheet.

In simple settings, one can take $S$ to be the thermodynamic entropy of isolated systems crossing the light-sheet. More generally, the definition of $S$ is subtle, because in field theory there are divergent contributions from vacuum entanglement across $\sigma_1$ and $\sigma_2$. Precise definitions of $S$ were found only recently, leading to rigorous proofs of two different field theory limits ($G\to 0$) of Eq.~(\ref{eq-genericbound}).  The proofs apply to free~\cite{BouCas14a,BouFis15b} and interacting~\cite{BouCas14b} scalar fields. Entropy bounds have also been verified~\cite{BouCas14b} or proven~\cite{KoeLei15} holographically for interacting gauge fields with a gravity dual.

Gravitational waves heat water, so they can be used to send information. In general, it is challenging to distinguish between gravitational waves and a curved spacetime. This can be done approximately in a setting where the wavelength of the radiation is small compared to other curvature radii in the geometry. A more rigorous notion of gravitational radiation is the ``Bondi news,'' which is defined in terms of an asymptotic expansion of the metric of asymptotically flat spacetimes~\cite{BMS,Sachs}. 

\begin{figure}[h]
	\centering
	\includegraphics[width=.45\columnwidth]{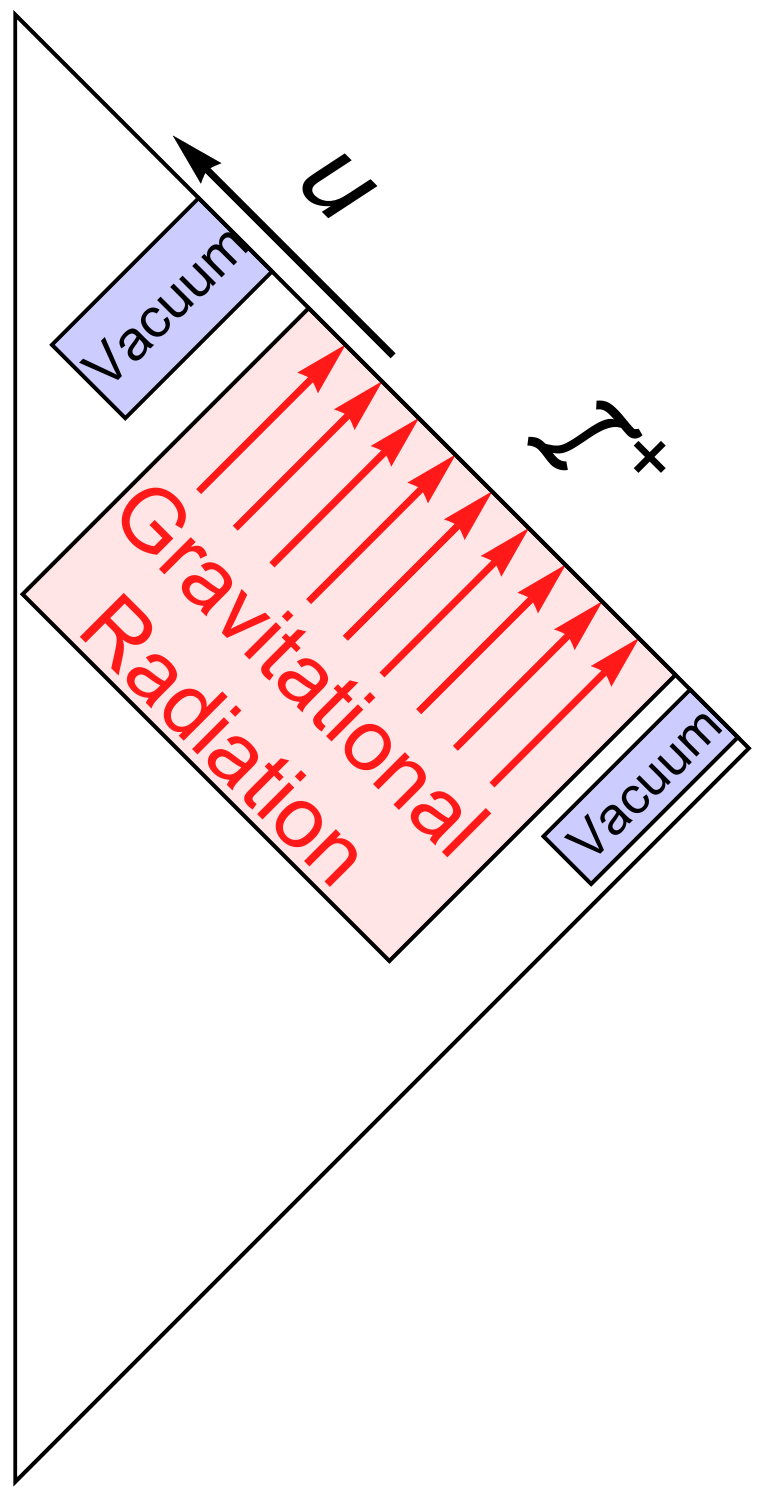}
	\caption{Penrose diagram of an asymptotic flat spacetime. Gravitational radiation (i.e., Bondi news) arrives on a bounded portion of ${\cal I}^+$ (red). The asymptotic regions before and after this burst (blue) are Riemann flat. The equivalence principle requires that observers with access only to the flat regions cannot extract classical information; however, an observer with access to the Bondi news can receive information (see Sec.~\ref{sec-soft}). We find in Sec.~\ref{sec-ic} that the asymptotic entropy bounds of Sec.~\ref{sec-bn} are consistent with these conclusions.}\label{fig:mink}
\end{figure}

The Bondi news corresponds to gravitational radiation that reaches distant regions (see Fig.~\ref{fig:mink}). It has been observed by monitoring test masses far from the source~\cite{LIGO1,LIGO2}. Its definition contains a rescaling by a factor of the radius, so that it remains finite as the radiation is diluted and weakened. Ultimately, it can be thought of as a spin-2 degree of freedom on future null infinity, ${\cal I}^+$. 

Recently, bounds on the entropy of arbitrary subregions of ${\cal I}^+$ were obtained as the limit of known bulk entropy bounds~\cite{Bou16}. These bounds constrain both the vacuum-subtracted entropy of states reduced to the subregion, and its derivatives as the subregion is varied. However, only nongravitational fields were treated rigorously. In this paper, we show how to incorporate gravitational radiation into the asymptotic entropy bounds. 

The bulk entropy bounds that formed the starting point of Ref.~\cite{Bou16} have been proven for certain fields~\cite{Wal11,BouCas14a,BouCas14b,BouFis15b,KoeLei15}. Unless there is a discontinuity in the asymptotic limit, we expect these proofs to apply to the asymptotic bounds as well. Explicit proofs have not yet been given for a spin-2 field, however. To be conservative, the asymptotic bounds on gravitational radiation should be regarded as a conjecture.

Therefore, we will perform a simple consistency check: we ask whether the bounds are compatible with the equivalence principle. We take this principle to be the statement that an empty, Riemann-flat spacetime region contains no classical information. (By this we mean a subset of Minkowski space, not of a Riemann-flat spacetime with nontrivial topology. In this paper, ``flat'' will always mean Riemann-flat and devoid of matter.) In particular, the classical information of the spacetime geometry is contained only in its Riemann curvature, and not, for example, in the choice of coordinates. 

The simplest setting is empty Minkowski space. In any subregion of ${\cal I}^+$, our upper bounds vanish, implying that the vacuum-subtracted entropy is nonpositive and independent of the subregion. (In particular, the upper bounds do not depend on a ``choice of vacuum'' of Minkowski space.) This is consistent with the equivalence principle, which tells us that no classical information is present. 

The asymptotic metric of Minkowski space, written in Bondi coordinates, is not uniquely fixed by fall-off conditions. One can freely choose the $1/r$ correction to the shape of spheres specified by setting the coordinates $u$ and $r$ to constants. (Note that this correction describes the shape of an embedded surface, whose location is determined by an arbitrary coordinate choice. Its shape is not indicative of any actual curvature of Minkowski space, which is manifestly Riemann-flat.) The freedom corresponds to a choice of a single real function $c(\Omega)$ of the coordinates on the sphere. 

Recently, this degeneracy in the choice of Bondi coordinates has been interpreted as a degeneracy of the actual vacuum state of Minkowski space~\cite{Str13,He:2014laa}. We take no position on the formal convenience of elevating a classical coordinate choice to a degeneracy of the vacuum. 

However, the equivalence principle rules out the possibility that a coordinate choice in Minkowski space has any measurable consequences. Therefore, $c(\Omega)$ must be unobservable. This is consistent with the fact that our bounds are insensitive to $c(\Omega)$ and vanish identically in Minkowski space.

We also consider a classical gravitational wavepacket with finite support, which arrives at ${\cal I}^+$ as Bondi news. In portions of ${\cal I}^+$ where the news has no support, our upper bounds vanish. This is consistent with the absence of classical information according to the equivalence principle: distant regions without gravitational radiation are Riemann-flat, so their geometry cannot be distinguished from Minkowski space. 

In Bondi coordinates, the Bondi news does change the function $c(\Omega)$, by an integral of the news~\cite{Str13,StrZhi14,He:2014laa}. Since the news can be measured, this integral can be measured; for example, it results in a permanent displacement of physical detectors. Thus, in a nonvacuum spacetime, $c(\Omega)$ is a coordinate choice only in that it can be picked freely {\em either} before {\em or} after the burst. The difference---the gravitational memory---is invariant and physical. The equivalence principle, and our bounds, constrain {\em how} the memory can be observed: namely, only by recording the news with physical detectors (which must be present during the burst). The memory cannot be measured by merely probing the asymptotic vacuum regions before and after the burst.

\paragraph*{Outline} In Sec.~\ref{sec-bn}, we review the derivation of asymptotic entropy bounds of Ref.~\cite{Bou16} (Sec.~\ref{sec-bounds}), and we show that they respond to gravitational radiation through the square of the Bondi news (Sec.~\ref{sec-shear}).

In Sec.~\ref{sec-soft}, we discuss implications of the equivalence principle. In Sec.~\ref{sec-vacuum}, we consider the term $C_{AB}(\infty)$ that appears in the asymptotic (Bondi) metric of asymptotically flat spacetimes. This term can be nonvanishing even in Minkowski space and has been interpreted as labelling degenerate vacua~\cite{Str13,Ash81,He:2014laa}. Since it corresponds to a coordinate choice in Minkowski space, the equivalence principle demands that $C_{AB}$ be unobservable in any experiment. In Sec.~\ref{sec-gm}, we consider gravitational memory (the integral of Bondi news). The equivalence principle implies that the memory can only be measured by an observer or apparatus that has access to all the Bondi news that produces the memory. In Sec.~\ref{sec-sg}, we discuss ``soft'' gravitons and gravitational waves, by which we mean waves with long wavelength compared to some other time scale in the process that produces them. Given enough time, such excitations can be distinguished from the vacuum and so their information content is unconstrained by the equivalence principle.

In Sec.~\ref{sec-ic}, we discuss implications of the entropy bounds of Sec.~\ref{sec-bounds}, in the same settings considered in Sec.~\ref{sec-soft}. In Minkowski space, there is no news, and all our upper bounds vanish. We also consider a classical probabilistic ensemble (i.e., a mixed state) of classical gravitational wave bursts. We find that our bounds permit an observer to distinguish between different classical messages if and only if the observer has access to the news. Thus the implications of our entropy bounds are consistent with the conclusions we draw from the equivalence principle in Sec.~\ref{sec-soft}. 

In Appendix~\ref{sec-krs}, we discuss an asymptotic entropy bound proposed by Kapec {\em et al.}~\cite{KapRac16}. We focus on the case of empty Minkowski space. Whether this bound differs from (a special case of) ours depends on the definition of the entropy, which was not fully specified in Ref.~\cite{KapRac16}. We argue that consistency with the equivalence principle requires a choice under which the bounds agree. We clarify that the extra term in the upper bound of Ref.~\cite{KapRac16} originates from a difference in how the relevant null surfaces are constructed before the asymptotic limit is taken. 

In Appendix~\ref{sec-singlePacket}, we apply the bounds of Sec.~\ref{sec-bounds} to a single graviton wavepacket. This case is not obviously constrained by the equivalence principle and so lies outside the main line of argument pursued here. We find that our bounds have implications similar to those derived for the classical Bondi news in Sec.~\ref{sec-ic}.

\section{Asymptotic Entropy Bounds and Bondi News}
\label{sec-bn}

In Ref.~\cite{Bou16}, entropy bounds were applied to a distant planar light-sheet. The bounds can be expressed in terms of the stress tensor of matter crossing the light-sheet, and the square of the shear of the light-sheet. It was shown that the matter contribution is independent of the orientation of the light-sheet in the asymptotic limit. However this was not proven for the contribution from the shear. Here we fill this gap by demonstrating that the shear term contributes to the upper bounds as the square of the Bondi news. Thus it is associated with gravitational radiation reaching the boundary. In particular, this implies that the asymptotic bounds of Ref.~\cite{Bou16} are fully independent of the orientation of the light-sheets used to derive them.

\subsection{Asymptotic Entropy Bounds}
\label{sec-bounds}

In this subsection we briefly review the derivation and formulation of the asymptotic entropy bounds of Ref.~\cite{Bou16}.   Expectation value brackets are left implicit throughout.

We consider entropy bounds~\cite{CEB1,FlaMar99,BouFis15a} in the general form of Eq.~(\ref{eq-genericbound}).  In the weak-gravity limit, Newton's constant $G$ is taken to become small, and a light-sheet is chosen that consists of initially parallel light-rays ($\theta_0=0$). An example of this is a null plane $t-z=\mbox{const}$ in Minkowski space. The effects of matter on the light-sheet are computed to leading nontrivial order in $G$, from the focussing equation~\cite{Wald}
\begin{equation}
-\frac{d\theta}{dw} = 8\pi G \,T_{ab} k^a k^b + \varsigma_{ab}\varsigma^{ab}~.
\label{eq-raych}
\end{equation}
Here $T_{ab}$ is the matter stress tensor, $k^a$ is the tangent vector to the light-rays that comprise the light-sheet, $w$ is an affine parameter and $\varsigma$ is the shear (defined by Eq.~\ref{eq-sheardef}). The expansion $\theta$ is the logarithmic derivative of the area of a cross-section spanned by infinitesimally nearby light-rays. 

By Eq.~(\ref{eq-genericbound}), the upper bound is given by the total area loss between two cross-sections of the light-sheet. It can be computed by integrating Eq.~(\ref{eq-raych}) twice along the light-rays, and then across the transverse directions. If the shear scales as $G^{1/2}$, the area loss will scale as $G$, so Newton's constant drops out in Eq.~(\ref{eq-genericbound}). The resulting bound involves only Planck's constant $\hbar$, so it can be viewed as a pure field theory statement. 

Near the boundary of an asymptotically flat spacetime, the matter stress tensor falls off as $r^{-2}$ and the shear associated with gravitational radiation falls of as $r^{-1}$, so the above argument can be carried out at finite $G$, as an expansion in $G/r^2$. In particular, one can work on a Minkowski background,
\begin{equation}
ds^2 = -du^2 - 2 du\, dr + r^2 d\Omega^2~,
\end{equation}
and compute area differences at order $G/r^2$, by integrating the focussing equation (\ref{eq-raych}).

Keeping the radiation under consideration fixed, the area of the radiation front increases in the asymptotic limit as the local stress tensor decreases. It is convenient to rescale both~\cite{Bou16}, and formulate asymptotic entropy bounds directly in terms of finite quantities on ${\cal I}^+$. The asymptotic energy flux is the energy arriving on ${\cal I}^+$ per unit advanced time and unit solid angle:\footnote{We will generally refer to boundary versions of bulk quantities by adding a hat.}
\begin{equation}
\hat{\cal T} = \hat T_{uu} + \hat\varsigma_{ab}\hat\varsigma^{ab}~,
\label{eq-oldhat}
\end{equation}
The first term is the energy flux of nongravitational radiation, 
\begin{equation}
\hat T_{uu} = \lim_{r\to\infty} r^2 T_{uu}~,
\end{equation}
The second term is set by the shear of the light-sheet and will be defined in Sec.~\ref{sec-shear}. It will be shown to correspond to the energy delivered by gravitational waves. 

In Ref.~\cite{Bou16}, the basic tool for deriving the asymptotic entropy bounds is the notion of a distant planar light-sheet. Let $p\in {\cal I}^+$ be a point at affine time $u_p$ and angle $\vartheta_p=\pi$. Let $H(u_p)$ be the boundary of the past of $p$:
\begin{equation}
H(u_p)\equiv \dot I^-(p)~,~~p\in {\cal I}^+ ~.
\label{eq-hdef}
\end{equation}
As discussed in Ref.~\cite{Bou16}, $H(u_p)$ is a null hypersurface. At $O(G/u_p^2)^0$, it is the null plane $t+z=u_p$ in Minkowski space, with affine parameter $w\equiv t-z$ and tangent vector $k^\mu=dx^\mu/dw$.  In the $\{u,r,\vartheta,\phi\}$ coordinates, $k^\mu$ has components
\begin{eqnarray} 
k^u
& = & \cos^2(\vartheta/2)
\label{eq-ku} \\
k^r 
& = & -(\cos\vartheta)/2
\label{eq-kr}\\
k^\vartheta
& = & \sin\vartheta\cos^2(\vartheta/2)/(u_p-u)
\label{eq-ktheta}\\
k^\phi
& = & 0~.
\label{eq-kphi}
\end{eqnarray}

In Ref.~\cite{Bou16} it was shown that a number of known weak-gravity entropy bounds apply on $H(u_p)$. Cuts on different $H(u_p)$ were identified for different $u_p$ by using the same function $u(\Omega)$ to define each cut; this function also defines a cut on ${\cal I}^+$. Bulk entropy bounds were applied to subregions defined by the cuts. The limit as $u_p\to\infty$ was taken and the bulk entropy bounds were re-expressed in terms of the asymptotic energy flux $\hat{\cal T}$. We will now list these results; see Ref.~\cite{Bou16} for details.

From the Quantum Null Energy Condition~\cite{BouFis15a,BouFis15b} (QNEC) on $H(u_p)$, one obtains the Boundary QNEC,
\begin{equation}
\frac{1}{\delta\Omega} \frac{d^2}{du^2} \hat S_\text{out}[\hat\sigma,\Omega] \leq \frac{2\pi}{\hbar}~ 
\hat{\cal T}~.
\label{eq-qnecscri0}
\end{equation}
Here $\delta\Omega$ is a small solid angle element near a null geodesic at angle $\Omega$ on ${\cal I}^+$. The second derivative is computed as this element is pushed to larger $u$, starting a given cut $\hat\sigma$ of ${\cal I}^+$. The limit as $\delta\Omega\to 0$ is implicit. The entropy $\hat S_\text{out}$ is the von Neumann entropy of the state of the subregion of ${\cal I}^+$ above the cut. That is,
\begin{equation}
\hat{S}_{\rm out} \equiv - \mathrm{tr}_{>\hat\sigma}\, \rho \log \rho~.
\end{equation}
The reduced state in the region above the cut $\hat\sigma$ is defined by
\begin{equation}
\rho = \mathrm{tr}_{<\hat\sigma}\, \rho_g~,
\end{equation}
where $\rho_g$ is the global state on ${\cal I}^+$. We need not include future timelike infinite since we assume that all matter decays to radiation at sufficiently late times. Note that all cuts of $\cal I^+$ have the same intrinsic and extrinsic geometry.  Therefore, divergent terms in $S_\text{out}$ drop out when differences are computed, or when derivatives are taken (also below). With the above definition, the QNEC has been proven for free scalar fields, and also for interacting gauge fields with a gravity dual~\cite{KoeLei15}.

From the differential, weak gravity Generalized Second Law (GSL) on $H(u_p)$~\cite{Bek72,Wal11}, or by integrating Eq.~(\ref{eq-qnecscri0}), one obtains the Boundary GSL in differential form
\begin{equation}
-\frac{1}{\delta\Omega} \frac{d}{du}\hat S_\text{out}[\hat\sigma;\Omega] \leq 
\frac{2\pi}{\hbar}\int_{\hat\sigma}^\infty du~ \hat{\cal T}~.
\label{eq-gslscri0}
\end{equation}
From the integrated weak-gravity GSL on $H(u_p)$, or by integrating Eq.~(\ref{eq-gslscri0}), one obtains the Boundary GSL in integral form,
\begin{equation}
\hat S_\text{out}[\hat\sigma_2]-\hat S_\text{out}[\infty]\leq
\frac{2\pi}{\hbar}\!\int_{\hat\sigma_2}^{\infty}\!\!\! d^2\Omega\, du\, 
[u-u_2(\Omega)]\, \hat{\cal T}~,
\label{eq-gslintscri0}
\end{equation}
where \(\hat S_\text{out}[\infty]\) is to be understood in a limiting sense.

Finally, from the Quantum Bousso Bound (QBB) ~\cite{BouCas14a,BouCas14b} on finite ``slabs'' of $H(u_p)$ one obtains a Boundary QBB,
\begin{equation}
\hat S_C[\hat\sigma_1,\hat\sigma_2]\leq \frac{2 \pi}{\hbar} \int_{\hat\sigma_2}^{\hat\sigma_1}\!\!\! d^2\Omega\, du\,  \hat g(u)\, \hat{\cal T}(u,\Omega)~.
\label{eq-qbbscri0}
\end{equation}
The weighting function $\hat g$ is different for free and interacting bulk fields~\cite{BouCas14a,BouCas14b}.  Since fields become free asymptotically, we expect that it is given by the free field expression $\hat g(u) = (u_1-u)(u-u_2)/(u_1-u_2)$. 

In Eq.~(\ref{eq-qbbscri0}), $\hat S_C$ is the vacuum-subtracted entropy~\cite{MarMin04,Cas08} of a finite affine interval on ${\cal I}^+$. It is defined directly on the finite portion of the light-sheet between $\hat \sigma_1$ and $\hat \sigma_2$, as the difference of two von Neumann entropies
\begin{equation}
\hat S_C[\hat\sigma_1,\hat\sigma_2] = -\mathrm{tr}\, \rho\log \rho + \mathrm{tr}\,\chi\log \chi~.
\label{eq-vacsub}
\end{equation}
Here the density operator $\rho$ is obtained from the global quantum state $\rho_g$ by tracing out the exterior of the region between $\hat\sigma_1$ and $\hat\sigma_2$; and $\chi$ is similarly obtained from the global vacuum state.\footnote{As discussed in the introduction, the equivalence principle requires that the reduced vacuum state is unique, so it implies that the definition of the vacuum-subtracted entropy is unambiguous. We return to this point in App.~\ref{sec-krs}.} The ultraviolet contributions from vacuum entanglement are the same in both reduced states, so they cancel out~\cite{MarMin04,Cas08,BouCas14a}. With this definition, the QBB has been proven both for free and interacting scalar fields. It has also been verified for gauge fields with gravity duals~\cite{BouCas14b}.

For free theories, the algebra of operators factorizes over the null geodesics that generate the light-sheet~\cite{Wal11}. We expect that this case applies to ${\cal I}^+$. Then the von Neumann entropy of the vacuum state restricted to the semi-infinite region above a cut $\hat\sigma$ is independent of the cut. Therefore, we have
\begin{equation}
\hat S_\text{out} [\hat\sigma_2]-\hat S_\text{out}[\hat\sigma_1] = 
\hat S_C[\hat\sigma_2]-\hat S_C[\hat\sigma_1]~,
\end{equation}
where $\hat S_C[\hat \sigma]$ is now computed on the semi-infinite regions above $\hat\sigma_1$ and $\hat\sigma_2$. Thus we can also express other bounds, Eqs.~(\ref{eq-qnecscri0}) and (\ref{eq-gslscri0}), in terms of derivatives and differences of the manifestly finite quantity $\hat S_C$, instead of $\hat S_\text{out}$. 

In particular, we can write the integrated Boundary GSL,  Eq.~(\ref{eq-gslintscri0}), as
\begin{equation}
\hat S_C[\hat\sigma_2]\leq
\frac{2\pi}{\hbar}\!\int_{\hat\sigma_2}^{\infty}\!\!\! d^2\Omega\, du\, 
[u-u_2(\Omega)]\, \hat{\cal T}~.
\label{eq-gslintscric}
\end{equation}
We have used the fact that the reduced density matrix of any physical state above a cut at sufficiently large \(u\) is that of the vacuum restricted to the same region, and thus  \(\hat S_C[\infty] = 0\). We will use this form of the integrated Boundary GSL in Sec.~\ref{sec-ic}.

\subsection{Bondi News as Shear on Distant Light-Sheets}
\label{sec-shear}

Let $\varsigma_{ab}$ be the shear tensor on $H(u_p)$, defined as the tracefree part of the extrinsic curvature: 
\begin{equation}
\varsigma_{ab} = B_{ab}-\frac{1}{2}\theta q_{ab}~,\label{eq-sheardef}
\end{equation}
where $B_{ab} = q_a^{~c} q_b^{~d} \nabla_c k_d$, and $q_{ab}$ is the metric on the cuts $w=$ const. One could choose different cuts, but some foliation of $H(u_p)$ into cuts has to be chosen in order to discuss the evolution of the shear. The shear tensor has only transverse components, so its information is fully captured by the lower-dimensional tensor
\begin{equation}
\varsigma_{\bar{A}\bar{B}} \equiv \varsigma_{ab} e^a_{~\bar{A}} e^b_{~\bar{B}}~.
\label{eq-ldt}
\end{equation}
The $D-2$ orthonormal vectors $e^a_{~\bar{A}}$ are tangent to the
cut. Below we will denote any projection with the $e^a_{~\bar{A}}$ by capital indices placed on higher-dimensional tensors. 

The evolution equation for the shear is~\cite{Wald,Poisson}
\begin{equation}
\frac{d}{dw}\varsigma_{\bar{A}\bar{B}} = W_{\bar{A}\bar{B}}- \theta\, \varsigma_{\bar{A}\bar{B}}~,
\label{eq-ww}
\end{equation}
where
\begin{equation}
W_{\bar{A}\bar{B}} = -\,C_{abcd} e^a_{~\bar{A}} k^b e^c_{~\bar{B}} k^d\equiv -\,C_{\bar{A}b\bar{B}d} k^b k^d
\end{equation}
and $C_{abcd}$ is the Weyl tensor. We now recall that at fixed $(u,\Omega)$ there is no difference between expansions in inverse powers of $u_p$ and $r$, since~\cite{Bou16}
\begin{equation}
r = \frac{u_p-u}{2\cos^2 (\vartheta/2)}~.
\label{eq-rutheta}
\end{equation}
The asymptotic behavior of the Weyl tensor is ~\cite{FlaNic15}
\begin{eqnarray} 
C_{\bar{A}u\bar{B}\vartheta} & \sim & O(r^{-1})~, \\
C_{\bar{A}u\bar{B}r}& \sim & O(r^{-3})~,\\
C_{\bar{A}r\bar{B}r}& \sim & O(r^{-4})~,\\
C_{\bar{A}r\bar{B}\vartheta}& \sim & O(r^{-3})~,\\
C_{\bar{A}\vartheta \bar{B}\vartheta}& \sim & O(r^{-1})~;
\end{eqnarray} 
and from Eqs.~(\ref{eq-ktheta}) and (\ref{eq-kphi}) 
\begin{equation}
k^\vartheta\sim O(r^{-1})~,~~k^\phi=0~.
\end{equation}
Hence we have
\begin{equation}
W_{\bar{A}\bar{B}} = -\,C_{\bar{A}u\bar{B}u} (k^u)^2 + O(u_p^{-2})~.
\end{equation}
These Weyl components are related to the Bondi news, $N_{AB}$~\cite{FlaNic15}:
\begin{equation}
C_{\bar{A}u\bar{B}u} = -\frac{1}{2r}\frac{d}{du} N_{AB} + O(r^{-2})~.
\end{equation}
We have introduced an unbarred basis defined by \(e^{a}_{~A} = r e^{a}_{~\bar{A}}\), with the feature that in this basis boundary quantities such as \(C_{AB}\) and \(N_{AB}\) are independent of \(r\). Unbarred capital indices will be raised and lowered with the unit two sphere metric, $h_{AB}$.

Since the expansion of \(H(u_{p})\) is of order $G/r^2$, the \(\theta\,\varsigma_{\bar{A}\bar{B}}\) term in Eq.~(\ref{eq-ww}) is always subleading in our analysis. Because the Bondi news and the shear of $H(u_p)$ both vanish in the far future, Eq.~(\ref{eq-ww}) implies
\begin{equation}
\varsigma_{\bar{A}\bar{B}} = \frac{1}{2r} N_{AB} \cos^2(\vartheta/2)+ O(r^{-2})~,
\label{eq-result}
\end{equation}
where we have used $d^2u/dw^2 \sim O(r^{-1})$. 

On the other hand, the ``boundary shear tensor'' appearing in Eq.~(\ref{eq-oldhat}) was defined in Ref.~\cite{Bou16} as
\begin{equation}
\hat\varsigma_{ab}(u,\vartheta,\phi) \equiv \frac{1}{\sqrt{8\pi G}}\lim_{r\to\infty} r\, \frac{\varsigma_{ab} (u,r,\vartheta,\phi)}{\cos^2(\vartheta/2)}~.
\label{eq-hatsigmadef}
\end{equation}

Comparing the previous two equations and using Eq.~(\ref{eq-ldt}), we recognize that the boundary shear is the Bondi news, up to an $O(1)$ rescaling:\footnote{In the Newman-Penrose formalism, the Bondi news is commonly identified with the \(u\)-derivative of the shear of the family of \emph{outgoing} null congruences specified by \(u=\) const \cite{Newman:1966ub}. Here we relate the news to the shear of \emph{ingoing} null congruences.}
\begin{equation}
\hat\varsigma_{AB}= \frac{N_{AB}}{\sqrt{32\pi G}}~.
\label{eq-sigmanews}
\end{equation}
The factor of $G^{-1/2}$ ensures that $\hat\varsigma^2$ has the dimension of an energy flux. 

Returning to the definition of the total asymptotic energy flux, Eq.~(\ref{eq-oldhat}), we can now write ${\cal T}$ in terms of the Bondi news:
\begin{equation}
\hat{\cal T} = \hat T_{uu} + \frac{1}{32\pi G} N_{AB} N^{AB}~,
\label{eq-newhat}
\end{equation}
Note that the definition of the boundary shear $\hat\varsigma_{AB}$ was tied to a family of null planes $H(u_p)$ whose orientation picks out a special point on the sphere. Since the Bondi news admits an independent definition that does not require us to pick such a point, it follows that the asymptotic bounds derived in Ref.~\cite{Bou16} are independent of the orientation of the $H(u_p)$.

In the remainder of this paper, we will specialize to the case where all outgoing radiation is gravitational. Then $\hat T_{uu}=0$ and $\hat{\cal T} = N_{AB} N^{AB}/32 \pi G$.  We see that the square of the Bondi news controls the entropy flux of gravitational radiation.

\section{Implications of the Equivalence Principle}
\label{sec-soft}

In this section, we consider classical aspects of gravitational radiation. We derive consequences of the equivalence principle: the hypothesis that no subset of Minkowski space contains any measurable classical information. Since we use the notion of classical information throughout this and the following sections, we begin with a simple example of such information and its description, in Sec.~\ref{sec-classical}.

It is possible to find nonvacuum {\em quantum\/} states whose effective stress tensor (analogous to Eq.~\eqref{eq-newhat}) vanishes in a bounded region. The geometry in this region could be Riemann-flat, yet the region could contain quantum information. Here we only assume the absence of classical information in Minkowski space. In particular we assume that no observable is associated with a coordinate choice in Minkowski space.

The geometry of Minkowski space is trivial, but of course the coordinates are arbitrary. So the matrix of metric components can take many different forms, both generally and in the asymptotic region. Restricting to Bondi coordinates does not fully fix this ambiguity. The equivalence principle implies that any parameters of the Bondi metric that are not unique in Minkowski space must be unobservable, or else that parameter would constitute measurable information. There is no $\hbar$ in the metric of Minkowski space in any Bondi gauge, so the corresponding coordinate information would be classical information. 

This includes in particular a parameter $C_{AB}$ (defined below) that has been interpreted~\cite{Ash81,Str13,He:2014laa} as labelling degenerate vacua (Sec.~\ref{sec-vacuum}). Indeed, no observation of this parameter has yet been made, and we are not aware of a proposal for how it could be measured.

A key consequence of the equivalence principle is that the gravitational memory created by Bondi news can be measured only by recording the news. It cannot be measured by probing the vacuum before and after the news (Sec.~\ref{sec-gm}). Finally, we note that the equivalence principle does not preclude soft gravitational radiation from carrying information, if ``soft'' is understood in the physically relevant sense of a small expansion parameter (Sec.~\ref{sec-sg}).

These conclusions are in harmony with our findings in Sec.~\ref{sec-ic}, where we apply the bounds of Sec.~\ref{sec-bounds} to constrain the information content of gravitons and of the vacuum.

\subsection{Classical Information} 
\label{sec-classical}

A simple example is a classical $n$-bit message written by Alice and delivered to Bob, say as a sequence of red and blue balls shot across space. Before Bob looks at the balls, he is ignorant of their state. Thus he can describe it as a density operator in a $2^n$ dimensional Hilbert space, which is diagonal in the $\{$red, blue$\}^n$ basis, with equal probability $2^{-n}$ for each possible message. The Shannon and von Neumann entropies are both 
\begin{equation}
-\sum_{i=1}^{2^n} p_i \log p_i = -{\rm Tr} \rho\log\rho=n\log 2~.
\end{equation}
This is an incoherent superposition, or classical probabilistic ensemble (not to be confused with a coherent quantum superposition of ball sequences). 

By looking at the balls, Bob learns Alice's message. Alice cannot send Bob more information than the maximum entropy of the system that carries the message. Since we can express Bob's initial ignorance as a density operator, quantum entropy bounds limit classical communication, as a special case.

Of course, the full quantum Hilbert space is much larger due to the internal degrees of freedom of the balls. And even in the tiny subfactor spanned by $\{$red, blue$\}^n$, more general states are possible at the quantum level, which are not product states of the individual balls. 

But for classical messages represented by a quantum density operator $\rho_i$, the ensemble interpretation~\cite{MikeIke} implies that the full density operator can be written as
\begin{equation}
\rho = \sum_{i=1}^{2^n} \rho_i~.
\end{equation}
Since the $\rho_i$ are classically distinguishable---and therefore mutually orthogonal---states, there is an irreducible uncertainty in the von Neumann entropy: the entropy cannot be parametrically less than the classical value, $n\log 2$. At the field theory level, this will remain true for the vacuum-subtracted von Neumann entropy: it must be parametrically at least $n\log 2$ (assuming the region contains all balls), since the vacuum entanglement is an ultraviolet quantum property shared by all the classical states.

In this paper we often consider the equivalence principle: the statement that Minkowski space, and any subset of it, contain no classical information. It is worth reflecting on what it would mean if empty Riemann flat space did contain measurable information. In that case it could be used by Alice to communicate a message to Bob.

To be concrete, consider an arbitrarily large patch of flat space (say, the interior of a falling elevator, or a large void in our universe). For it to contain information in an operationally meaningful sense, Alice would have to be capable of ``preparing'' this region, perhaps by sending a certain sequence of gravitational waves through it. Later, long after those waves have left the region and it is again empty and Riemann-flat, Bob would have to be capable of reading out the message that Alice ``left behind'', by examining only this patch. 

Specifically, if $c(\Omega)$ was observable, then independent observers with access only to the flat space region, would all come to the same conclusion as to which coordinates should be used to label its spacetime points. More precisely, up to corrections subleading in $1/r$, such observers would uniquely identify topological spheres on which the Bondi coordinates $u$ and $r$ must be constant, thus partially fixing the chart. This would indeed be a textbook violation of the equivalence principle. 

\subsection{Empty Space Has No Classical Information}
\label{sec-vacuum}

Let us consider the asymptotic metric of an asymptotically flat spacetime, in standard retarded Bondi coordinates (see, e.g.,~\cite{BMS,Sachs,Ash81,Str13,StrZhi14,FlaNic15}): 
\begin{eqnarray} 
ds^2 & = & -\left(1-\frac{2m_B(u,\Omega)}{r}\right) du^2 - 2\, du\, dr \nonumber \\ 
& + & r^2 \left(h_{AB}+\frac{C_{AB}(u,\Omega)}{r}\right) \times \nonumber\\
 & & ~~\times \left(d\vartheta^A+\frac{D_CC^{AC}}{2r^2} du\right)
\left(d\vartheta^B+\frac{D_CC^{CB}}{2r^2} du\right) \nonumber \\ 
& + & \ldots
\label{eq-bondi}
\end{eqnarray} 
where $m_B$ is the Bondi mass aspect, and the ellipses indicate terms subleading in $r$.  Here, $C_{AB}(u,\Omega)$ appears as the $1/r$ correction to the round two-sphere metric $h_{AB}$. It satisfies $h^{AB} C_{AB}=0$ and $C_{AB}=C_{BA}$. The Bondi news is defined by
\begin{equation}
N_{AB}=\partial_u C_{AB}~.
\label{eq-newsdef}
\end{equation}

In Minkowski space, the news vanishes. However, the asymptotic metric of Minkowski space can be written in the form of Eq.~(\ref{eq-bondi}), with $m_B\equiv 0$ and any $u$-independent choice of a tracefree symmetric $C_{AB}(\Omega)$ satisfying 
\begin{equation}
C_{AB}=(2D_A D_B - h_{AB}D_C D^C)\, c(\Omega)
\label{eq-vacc}
\end{equation}
for some function $c$ on the sphere. But of course, the geometry is always the same, no matter how we label its points. There is no curvature of any kind, whatever value we choose for $c(\Omega)$. By the equivalence principle, this implies that $c$ and $C_{AB}$ cannot be measured.

$C_{AB}$ does transform nontrivially under large diffeomorphisms of the asymptotic metric~\cite{Str13,Barnich:2009se,Barnich:2010eb,FlaNic15}; indeed, this is one way to see that it is non-unique in Minkowski space.  Under a BMS supertranslation, $u\to u+f(\Omega)$, one has
\begin{equation}
C_{AB} \to C_{AB}+\left (2 D_A D_B - h_{AB}D^C D_C\right) f(\Omega)
\end{equation}
in regions where $N_{AB}=0$. This corresponds to a well-defined {\em change} in the shape of a large coordinate sphere at constant $u,r$. It affects all such spheres equally; for example $C_{AB}(\infty)$ and $C_{AB}(-\infty)$ will change by the same amount under a supertranslation.  
Of course, this does not imply that $C_{AB}$ is observable. A coordinate sphere is not a physical object but a collection of spacetime points. Its initial shape before the transformation is set by a coordinate choice.

The transformation properties of $C_{AB}$ under supertranslations have been interpreted as an infinite ``vacuum degeneracy'' of Minkowski space~\cite{Str13,Ash81,He:2014laa}. Each ``vacuum'' is labeled by the function $c(\Omega)$ in Eq.~(\ref{eq-vacc}). We conclude that the equivalence principle precludes any observable consequences of this degeneracy.\footnote{Note that the equivalence principle only precludes diffeomorphisms from transforming the classical vacuum into a physically distinct configuration. The equivalence principle does not imply that large diffeomorphisms always act trivially. When acting on an excited state, a supertranslation generically produces a distinct excited state, for example with a different relative timing of the Bondi news arriving at different angles.} (Refs.~\cite{Averin:2016hhm,Dvali:2015rea} give an argument that the vacua are indistinguishable starting from different assumptions.)

\subsection{Gravitational Memory}
\label{sec-gm}

In nonvacuum spacetimes, $C_{AB}$ need not be constant in $u$, and differences between $C_{AB}$ at different cuts are observable as ``gravitational memory.'' However, the value of $C_{AB}$ at any one cut (or its zero-mode) must be unobservable in any asymptotically flat spacetime, or else the equivalence principle would be violated in regions where no news arrives. We will now discuss this.

Suppose that some process (a binary inspiral, say) produces gravitational radiation, and that the corresponding Bondi news arrives entirely between the cuts $\hat\sigma_1$ and $\hat\sigma_2$ of ${\cal I}^+$.  The integral of the news along the null direction is called the gravitational memory produced by the process,
\begin{equation}
\Delta C_{AB}(\Omega) \equiv \int_{\hat \sigma_1}^{\hat \sigma_2} du \, N_{AB}(u,\Omega)
\label{eq-cabnab}
\end{equation}

By Eq.~(\ref{eq-cabnab}), the production of memory requires nonzero flux of radiation, $N_{AB}$. Hence memory production occurs only in excited states, not in the vacuum. For example, a graviton wavepacket can produce memory; but then the global state is not the vacuum, but a one-particle state. This qualitative fact continues to hold invariantly in the ``soft limit,'' as the wavepacket is taken to have arbitrarily large wavelength.

What is the physical manifestation of $\Delta C_{AB}$, or equivalently, how can it be measured? In Sec.~\ref{sec-shear}, we showed that $N_{AB}$ is proportional to the shear of a planar null congruence $H(u_p)$ near ${\cal I}^+$. Hence the gravitational memory is related to the integrated shear, i.e., the resulting strain of the congruence. The displacement vector $\eta^{\bar{A}}$ of two infinitesimally nearby null geodesics will change by 
\begin{equation}\label{eq-memoryEta}
\Delta \eta^{\bar{A}} =  \Delta C^{A}_{~\,B}\, \frac{\eta^{\bar{B}}}{2r}
\end{equation}
between $\hat\sigma_1$ and $\hat\sigma_2$. This can be measured by setting up (before $\hat\sigma_1$) a collection of physical, massless particles propagating along the null geodesics that constitute $H(u_p)$, and observing their transverse location on a screen that they hit after $\hat\sigma_2$. $\Delta C_{AB}$ can also be measured using an array of timelike detectors distributed over a large sphere. The displacement of any two detectors similarly suffers an overall change given by Eq.~\eqref{eq-memoryEta}.

In general, the memory captures only a small fraction of the information that arrives in distant regions: the integral of the Bondi news. It would certainly be nice to measure this component using gravitational wave detectors~\cite{Chr91,Tho91}. Such a measurement would not take infinite time, and it would not be conceptually distinct from any other measurement of the outgoing radiation. 

By Eq.~(\ref{eq-newsdef}) we can write the gravitational memory, Eq.~(\ref{eq-cabnab}), as a difference of the metric quantity $C_{AB}$ evaluated at the two cuts, 
\begin{equation}
\Delta C_{AB}(\Omega) = C_{AB}[\hat \sigma_2]-C_{AB}[\hat \sigma_1]~.
\label{eq-memory}
\end{equation} 
If $C_{AB}$ is interpreted as labelling a vacuum, the creation of gravitational memory by news could be described as a ``transition'' between two such vacua.  However, according to the equivalence principle this language is misleading, because $C_{AB}$ cannot be observed at a local cut. A ``vacuum'' in the above sense is a coordinate label that contains no physical information. 

Only the difference $\Delta C_{AB}$ is invariant (up to Lorentz transformations~\cite{FlaNic15}) and so can be observed.  $\Delta C_{AB}$ is nonzero only in global states which are {\em not} the vacuum, and it is fully determined by the integral of the Bondi news. So the function $C_{AB}(u,\Omega)$ contains no physical information beyond what is already in its $u$-derivative, the news $N_{AB}$. The observable memory, $\Delta C_{AB}$, captures a subset of the information in the news.  

By the equivalence principle, $\Delta C_{AB}$ can only be measured by an observer who has access to the entire region in which news arrives. For example, if physical test particles are introduced into the asymptotic region, and their initial position at $\hat\sigma_1$ is recorded, then the memory $\Delta C_{AB}$ can be measured at $\hat\sigma_2$ by observing the new location of these physical objects. This is an integrated measurement of the Bondi news, with the dynamics of the test masses doing the integration. 

Formally, it can be convenient to consider the ``zero mode'' of the news, 
\begin{equation}
C_{AB}(\Omega,\infty)-C_{AB}(\Omega,-\infty) \equiv  \int_{-\infty}^\infty du \, N_{AB}(\Omega,u)~.
\end{equation}
This quantity represents the total amount of memory produced in an asymptotically flat spacetime. As written, it is {\em not} observable, since no experiment began in the infinite past and will end in the infinite future. Fortunately, in any physical process or sequence of processes, the production of news will have a beginning and an end. So one can record the entire memory in a finite-duration experiment, corresponding to a sufficiently large finite range of integration.

To summarize both this and the previous subsection, the value of $C_{AB}$ at any one cut can be changed by a global change of coordinates. By the equivalence principle, $C_{AB}$ cannot be observed and contains no physical information. Therefore, in particular, we cannot measure the gravitational memory, $\Delta C_{AB}$ by observing $C_{AB}$ locally at $\hat\sigma_1$ and $\hat\sigma_2$ and computing the difference. Rather, physical test masses are essential for recording the news and integrating it to obtain $\Delta C_{AB}$ between the two cuts. If we forgot to introduce real test masses at $\hat\sigma_1$, we cannot look at empty space at $\hat\sigma_2$ and learn anything from it.

\subsection{Soft Gravitons}
\label{sec-sg}

A soft particle is an excitation of a massless field whose characteristic wavelength, or inverse frequency, is large compared to some dynamical timescale that otherwise characterizes a problem. For example, consider a binary system composed of neutron stars or black holes. They orbit each other with some frequency $\omega$, which varies slowly as they approach, until they eventually merge. The system will emit ``hard'' gravitational waves with frequency of order $\omega$. The overall duration of the inspiral process is much greater than $\omega^{-1}$; it is characterized by a second time scale $\tau\gg \omega^{-1}$. Or consider a black hole emitting Hawking radiation. The wavelength of the ``hard part'' of the radiation is of order the black hole radius, $\omega^{-1}\sim O(R)$, which changes slowly. Nonetheless, the overall process takes a much longer time, $\tau\sim O(R^3/G\hbar)$. 

Because the emission of ``hard'' radiation slowly transports gravitating energy from the center to distant regions, the gravitational field will vary not only with characteristic frequency $\omega$, but also over the timescale $\tau$. Therefore, signals with characteristic frequency as low as $\tau^{-1}$ are produced in the above processes. Such signals are referred to as ``soft''. (Often the term ``soft graviton'' is used, even when the signal is classical.)

This terminology is convenient when we wish to distinguish particles associated with different timescales in a given problem. Useful results can be obtained by expanding in ratios of such timescales~\cite{Wei65}. It can also be convenient to idealize soft particles by taking a $\tau\to\infty$ limit, for the purposes of making such expansions sharp. It is worth stressing, however, that infinite-duration experiments are not actually needed to produce and measure a soft particle. (If they were, soft particles would have no physical relevance.)  The larger time scale $\tau$ is necessarily finite in any physical process. 

Moreover, the production of observable radiation comes at a nonzero energy cost. If a soft graviton were added to the vacuum, one would obtain an excited state orthogonal to the vacuum, not a new vacuum. This is a qualitative statement, and independent of $\tau$. Thus, there is no fundamental difference between soft particles and any other form of radiation that arrives in distant regions. 

Correspondingly, when we apply the boundary entropy bounds of Sec.~\ref{sec-bounds} in Sec.~\ref{sec-ic}, all Bondi news can be treated on the same footing. For example, if the interval under consideration in Eq.~(\ref{eq-gslintscric}) or Eq.~(\ref{eq-qbbscri0}) is large enough to contain a news wavepacket (hard or soft), we will find that this graviton will contribute to the energy side, and generically also to the entropy side of the inequality.

\section{Entropy Bounds on Gravitational Wave Bursts and the Vacuum}
\label{sec-ic}

In this section, we compute the upper bounds of Sec.~\ref{sec-bounds} in simple asymptotically flat spacetimes: Minkowski space, and a burst of Bondi news that creates gravitational memory. We show that the upper bounds are consistent with constraints derived in the previous section from the equivalence principle.

\subsection{Vacuum}
\label{sec-vacb}

Let us apply the bounds of Sec.~\ref{sec-bounds} to empty Minkowski space: the Boundary QNEC, Eq.~(\ref{eq-qnecscri0}); the Boundary GSL in integrated and differential form, Eqs.~(\ref{eq-gslscri0}) and (\ref{eq-gslintscric}); and the Boundary QBB, Eq.~(\ref{eq-qbbscri0}). All of these bounds are linear in the boundary stress tensor $\hat{\cal T}$, i.e., quadratic in the Bondi news. Since $\hat{\cal T}=0$ in Minkowski space, the upper bounds all vanish. 

The Boundary QBB implies that the vacuum-subtracted entropy is nonpositive in any finite subregion of ${\cal I}^+$. The Boundary GSL implies that it is nonpositive for any semi-infinite region above a cut, and independent of the choice of region or deformations of the cut. The Boundary QFC implies (redundantly with the above) that the second derivative under deformations also vanishes. 

These upper bounds are consistent with all implications of the equivalence principle described in the previous section: no subset of Minkowski space contains any classical information. Moreover, both the bounds and the equivalence principle are consistent with the simplest possibility for the quantum description of Minkowski space: that the ground state is unique, and that the vacuum-subtracted entropy precisely vanishes on any subregion of ${\cal I}^+$. 

\subsection{Classical Bondi News}
\label{sec-state}

For simplicity, we will consider a single wave packet of gravitational radiation, of characteristic wavelength $\lambda$ in the $u$-direction. The wave packet is roughly centered on $u=0$ and delocalized on the sphere. The wave packet can be used to send a message to an observer at ${\cal I}^+$, for example by encoding it in its polarization, its shape, its direction (the angle $\Omega$ at which it arrives), or the time of arrival, within a finite discrete set of ${\cal N}$ possible choices. 

For concreteness, let us encode the information in the energy of the wavepacket. We take the energy to be of order $E$ for any message, but with a grading into ${\cal N}$ different values. A single graviton has energy of order $\hbar/\lambda$. Since we wish to work in the classical regime, the grading must be much coarser than that, so the number of distinct classical states will satisfy
\begin{equation}
{\cal N} \ll \frac{E\lambda}{\hbar}~,
\label{eq-nel}
\end{equation}
We assume that any of the distinct classical signals arrives with equal probability $1/{\cal N}$. The classical Shannon entropy is thus $\log {\cal N}$.

If we apply the Boundary QBB, Eq.~(\ref{eq-qbbscri0}), or the Boundary GSL, Eq.~(\ref{eq-gslintscric}), to the region occupied by the wavepacket, we obtain 
\begin{equation}
\hat S_C \lesssim \frac {E\lambda}{\hbar}
\label{eq-luck}
\end{equation}
This is consistent: in our example, the vacuum-subtracted entropy need not be much greater than the Shannon entropy $\log {\cal N}$, which is much smaller than ${\cal N}$ and hence, by Eq.~(\ref{eq-nel}), much smaller than the upper bound. Thus, the asymptotic bounds of Sec.~\ref{sec-bounds} easily accommodate the classical information contained in the Bondi news.

On the other hand, if we apply the same bounds to a region that fails to overlap with the wavepacket, then the upper bound vanishes:
\begin{equation}
\hat S_C \leq 0~.
\label{eq-noluck}
\end{equation}
This is consistent with the absence of classical information in asymptotic regions that do not contain news, as required by the equivalence principle. 

In particular, the bounds are consistent with our conclusion in Sec.~\ref{sec-soft} that gravitational memory can only be measured by an observer who has access to the news that creates the memory. In our present example, the news is featureless but for its overall energy. So its integral, the memory, contains the same amount of information as the news, $\log {\cal N}$. [We have ${\cal T}\sim N_{AB}N^{AB}/G\sim E/\lambda$, so the memory will be of order $\Delta C_{AB}\sim N_{AB}\lambda\sim (GE\lambda)^{1/2}$.] By Eq.~(\ref{eq-noluck}), this information is unavailable to an observer who cannot access the news.

\appendix
\section{KRS Bound}
\label{sec-krs}

Ultimately the equivalence principle is a hypothesis, supported by a certain amount of evidence. Indeed, Strominger~\cite{Str13} (building on earlier work of Ashtekar~\cite{Ash81} and others) has argued that the vacuum of asymptotically flat spacetimes, Minkowski space, is infinitely degenerate, i.e., that it corresponds to an infinite number of distinct quantum states labeled by the quantity $C_{AB}$ in Eq.~(\ref{eq-bondi}). 

If these states could be distinguished by any observation, empty space would contain an infinite amount of information. This would constitute a violation of the equivalence principle in its usual, classical sense: in a basis where Minkowski-like states are labeled by $C_{AB}$, they can be naturally identified with a choice of coordinates, so coordinates would be measurable. Kapec, Raclariu, and Strominger~\cite{KapRac16} (KRS) recently proposed an entropy bound that contains an extra term (denoted $X^{\rm KRS}$ below), designed to account for this possibility.

A precise definition of the relevant entropy was not yet given in Ref.~\cite{KapRac16}. More importantly, no measurement protocol has been suggested for extracting the information contained in empty space. Such a measurement would rule out the equivalence principle experimentally. Conversely, absent experimental evidence to the contrary, we would argue that the equivalence principle should be retained: we should not consider Minkowski space written with different coordinate parameters $C_{AB}$ to be physically distinct spacetimes.

In order to facilitate further study, we will summarize our understanding of the differences between our bounds and the KRS bound. We will offer a geometric interpretation of the extra term $X^{\rm KRS}$. We will explain why it appears in their derivation of an asymptotic bound but not in ours. We will also describe how the presence of this term conflicts with the equivalence principle.

KRS considered the asymptotic limit of bulk null hypersurfaces with approximately spherical cross-sections. This simplifies the approach to ${\cal I}^+$ in spherical Bondi coordinates, compared to our use of planar light-sheets. Unlike the planar null surfaces $H(u_p)$ used above, however, existing bulk entropy bounds become divergent and hence trivial in the asymptotic limit of spherical null surfaces. Hence they cannot be used as a starting point if one wishes to work with spherical cross-sections. A new subtraction method was proposed to cancel the divergence~\cite{KapRac16}. The KRS bound is
\begin{equation}
S_0^{\rm KRS}[\hat\sigma_2] \leq \Delta K[\hat\sigma_2]  + X^{\rm KRS}[\hat\sigma_2;C_{AB}(\infty)]
\label{eq-krs}
\end{equation}
where $u_2(\Omega)$ defines the position of the cut.\footnote{In the notation of Ref.~\cite{KapRac16}, our $\hat\sigma_2$ is their $\Sigma$; our $\Delta K$ is $-A^{\Sigma}_F/4G\hbar$; and our $X^{\rm KRS}$ is $(-A^{\Sigma}_0+ A^{\Sigma}_F)/4G\hbar$.} A full definition of $S_0^{\rm KRS}$ was left to future work, but we will argue below that the choice is tightly constrained by coordinate invariance.

Let us first consider the r.h.s.\ of Eq.~(\ref{eq-krs}). The first term is given by
\begin{equation}
\Delta K[\hat\sigma_2]\equiv \frac{2\pi}{\hbar}\!\int_{\hat\sigma_2}^{\infty}\!\!\! d^2\Omega\, du\, 
[u-u_2(\Omega)]\, \hat{\cal T}~.
\end{equation}

This is precisely the r.h.s.\ of our integrated Boundary GSL, Eq.~(\ref{eq-gslintscric}):
\begin{equation}
\hat S_C[\hat\sigma_2]\leq \Delta K[\hat\sigma_2]~.
\label{eq-gslbig}
\end{equation}

However the r.h.s.\ of the KRS bound contains the extra term
\begin{eqnarray} 
X^{\rm KRS} & \equiv & -\frac{1}{8G \hbar} \int d^2\Omega~ D_A u_2(\Omega) ~ D_B \bar C^{AB}
\label{eq-x1} \\
& = & \frac{1}{8G \hbar} \int d^2\Omega~ u_2(\Omega)~ D_AD_B  \bar C^{AB} \label{eq-x2}
\end{eqnarray} 
where
\begin{equation}
\bar C^{AB} \equiv C^{AB}(\infty)-C_{0}^{AB}~.
\label{eq-fid}
\end{equation}
Here $C_0^{AB}$ refers to a fiducial choice of Bondi coordinates (or of a ``late-time vacuum'' in the sense of Ref.~\cite{Str13}) at $u\to\infty$, whereas $C^{AB}$ refers to the ``actual'' Bondi coordinates (or ``late-time vacuum'') that will be attained as $u\to\infty$.  Because $D_A D_B \bar C^{AB}$ is a total derivative, its average on the cut $\hat\sigma_2$ vanishes, so unless it vanishes identically, it will have indefinite sign on the sphere. It also follows that $X^{\rm KRS}=0$ if $u_2=\mbox{const}$, so the extra term only contributes if the cut has nontrivial angular dependence in the chosen coordinates.

In the bulk, $D_A D_B \bar C^{AB}$ arises geometrically from a nonvanishing expansion of the null hypersurfaces at late times, which remains after the KRS regularization. Namely, the null expansion orthogonal to a surface of constant $u,r$ in Minkowski space in the metric of Eq.~(\ref{eq-bondi}) is
\begin{equation}
\theta[C_{AB}] = - \frac{1}{r} -\frac{1}{2r^2}D^A D^B C_{AB}~,
\end{equation}
so the difference between two choices $C^{AB}$, $C^{AB}_0$ yields
\begin{equation}
\bar\theta(\Omega)=-\frac{1}{2r^2}D^A D^B \bar C_{AB}~.
\end{equation}
Substituting this result in Eq.~(\ref{eq-x2}), the term $X^{\rm KRS}$ can thus be understood as an extra area difference accumulated due to a nonzero regulated expansion $\bar\theta$ of the KRS null surface at late times. 

The extra term $X^{\rm KRS}$ was motivated in Ref.~\cite{KapRac16} by covariance of their geometric construction under BMS transformations, so it is worth explaining its absence in Eq.~(\ref{eq-gslbig}) and our other bounds. KRS consider a coordinate sphere at fixed $u,r$ at late times, and construct a null hypersurface orthogonal to it. BMS transformations act nontrivially by deforming the geometry of this coordinate sphere and changing its null expansion as a function of angle. This change propagates along the entire null hypersurface and leads to an extra area difference $X^{\rm KRS}$ as described above.

A bulk BMS supertranslation of a given late-time cross-section of the null plane $H(u_p)$ would yield a similar term. The null surface $\tilde H(u_p)$ orthogonal to the new cross-section would be neither a light-sheet nor a causal horizon, because the expansion at late times has the wrong sign on some generators. From this perspective, the KRS conjecture involves a modification of the nonexpansion condition of the covariant entropy bound, such that the permitted range of the (regulated) expansion depends on the late-time Bondi frame.

However, with our definition of $H(u_p)$, BMS supertranslations do not act in this way. We defined $H(u_p)$ not in terms of a given bulk cross-section, but as the boundary of the past of a point $p$ on ${\cal I}^+$~\cite{Bou16}. BMS supertranslations can only move this point along the null geodesic generator on which it lies. The boundary of the past of any point on ${\cal I}^+$  has vanishing late-time expansion and is a causal horizon. Thus, supertranslations map the set of all $H(u_p)$ to itself. Therefore they have no effect when the limit as $u_p\to\infty$ is taken, and they leave no imprint in our asymptotic entropy bounds.

We now turn to the l.h.s.\ of Eq.~(\ref{eq-krs}). The indefinite sign of $D_A D_B C^{AB}(\infty)$ on the sphere constrains possible definitions of $S_0^{\rm KRS}$. It implies that $S_0^{\rm KRS}[\sigma_2]$ cannot be unique in Minkowski space, for any nonconstant cut $\sigma_2$. In particular, it is not possible for $S_0^{\rm KRS}$ to always vanish for arbitrary subregions of the boundary of Minkowski space regardless of the choice of coordinates.

To see this, choose asymptotic coordinates such that $\bar C_{AB}=\beta \tilde C_{AB}$ where $\tilde C_{AB}$ is nonvanishing and satisfies Eq.~(\ref{eq-vacc}), and \(\beta\) is a constant. By Eq.~(\ref{eq-x1}), $X^{\rm KRS}$ is linear in $\beta$ so it can be made negative and arbitrarily large in magnitude by an appropriate choice of $\beta$. This would violate the KRS bound so $S_0^{\rm KRS}[\sigma_2]$ must depend on $\bar C_{AB}$. 

The above considerations also imply that $S_0^{\rm KRS}$ cannot be bounded from below by the Shannon entropy---not even approximately---in the case where classical Bondi news is present.

Indeed, KRS advocate that $S_0^{\rm KRS}$ should not be unique in Minkowski space. Rather it should contain a ``soft term'' that depends on $C^{AB}$ in some way, so that the KRS bound is satisfied independently of the choice of the ``reference vacuum'' $C_0^{AB}$. 

Here we note that the only definition consistent with the equivalence principle is
\begin{equation}
S_0^{\rm KRS}\equiv \hat S_C+X^{\rm KRS}~,
\label{eq-skrsdef}
\end{equation}
where $S_C$ has no dependence on $C_{AB}$. With this choice, the $X^{\rm KRS}$ terms would cancel, and thus, all dependence on $C_{AB}$ would drop out. Then Eq.~(\ref{eq-krs}) would reduce to Eq.~(\ref{eq-gslbig}). With any inequivalent definition, the physical content of Eq.~(\ref{eq-krs}) would depend on a coordinate choice.
 
This is because $X^{\rm KRS}$ depends on the quantity $\bar C_{AB}$ defined in Eq.~(\ref{eq-fid}). We have argued in Sec.~\ref{sec-vacuum} that $C_{AB}(\infty)$ can be changed by changing the coordinate choice. Therefore, neither $C_{AB}(\infty)$ nor its difference from a fiducial value, $\bar C_{AB}$, can be observable, if the equivalence principle is valid. [Note that the fiducial value $C^{AB}_0$ need {\em not} correspond to the value of $C_{AB}$ at any cut on ${\cal I}^+$. If it did, $\bar C_{AB}$ could be measured, and it would originate with physical radiation whose information content satisfies Eq.~(\ref{eq-gslbig}).]

In particular, if $S_0^{\rm KRS}$ could be constructed entirely from observable quantities, then Eq.~(\ref{eq-krs}) could be used to constrain $\bar C_{AB}$, thus making it accessible to observation. This would be a problem: $\bar C_{AB}$ {\em must} remain unobservable by the equivalence principle, because it corresponds to a coordinate choice in Minkowski space. 

In closing, we stress again that by the equivalence principle we mean the statement that empty Riemann-flat space contains no classical information. 
In Sec.~\ref{sec-ic} we showed that the bounds of Sec.~\ref{sec-bounds} are consistent with this principle. In this appendix we have argued that the KRS bound is not consistent with it, except for a particular choice of definition of entropy under which it would reduce to Eq.~(\ref{eq-gslintscric}). We make no claims about the compatibility of the KRS bound with any other formulation of the equivalence principle.

\section{Single Graviton Wavepacket}
\label{sec-singlePacket}

In this appendix, we study the implications of asymptotic bounds in a quantum setting; we will find that in some cases they restrict the entropy more strongly than the equivalence principle did for classical waves.

We consider a classical probabilistic ensemble of single graviton wave packets, of characteristic wavelength $\lambda$ in the $u$-direction. Like the classical gravitational wave of Sec.~\ref{sec-state}, the wave packets shall be roughly centered on $u=0$, and delocalized on the sphere. This is a global quantum state, defined on all of ${\cal I}^+$. 

Any such state is orthogonal to the vacuum. Here we shall take the global state $\rho_g$ to be a mixed state with global von Neumann entropy of order unity:
\begin{equation}
\hat S_g = - \mathrm{tr}\, \rho_g\log\rho_g \sim O(1)~.
\end{equation}
For example, $\rho_g$ could be an incoherent superposition of the graviton wavepacket in two different polarization states. Alice could encode a message about the weather in the choice of polarization, and Bob could decode this message if he is able to measure the polarization.

In the region occupied by the wave packet, we have
\begin{eqnarray} 
 N_{AB}N^{AB}  & \sim & O\left(\frac{l_P^2}{\lambda^2}\right) \label{eq-nn}~,\\
 \hat{\cal T}  & \sim & O\left(\frac{\hbar}{\lambda^2}\right) \label{eq-calt}~,\\
 N_{AB}  & \sim & O\left(\frac{l_p}{\lambda}\right) \label{eq-n}~,
\end{eqnarray} 
where expectation value brackets are left implicit.
The gravitational memory created by the wavepacket is
\begin{equation}
 \Delta C_{AB}^\infty   = \int_{-\infty}^\infty  N_{AB}\,  
du\approx\int_{-\lambda}^\lambda   N_{AB} \,du \sim  O(l_P)~, \label{eq-dci}
\end{equation} 
where
\begin{equation}
l_P\equiv \sqrt{G\hbar}
\end{equation}
is the Planck length. Note that the memory is independent of $\lambda$ and so remains finite as $\lambda$ is taken large.

\subsubsection{Boundary Quantum Bousso Bound}

The Boundary QBB, Eq.~(\ref{eq-qbbscri0}), bounds the entropy on finite portions of ${\cal I}^+$. This is particularly relevant to actual experiments. There are no experiments that started infinitely long ago and will complete an infinite time from now. When we measure something, we do it in finite time. 

Hence, we will consider an experiment of finite duration of order $T$. It will be convenient to center this time interval near $u=0$. Thus, we consider an observer who has access to the subregion
\begin{equation}
-T\lesssim u \lesssim T
\end{equation}
of ${\cal I}^+$ (or to the subregion of the asymptotic region defined by the same range, in Bondi coordinates). It will not be important whether the cuts $\hat\sigma_1$, $\hat\sigma_2$ are at constant $u$.

All observables that can be measured by this observer can be computed from the reduced density operator
\begin{equation}
\rho_T \equiv \mathrm{tr}_{\not T}\, \rho_g~.
\end{equation}
We must also consider the global vacuum state, restricted to the observation interval:
\begin{equation}
\chi_T=\mathrm{tr}_{\not T}\,  |0\rangle \langle 0|~.
\label{eq-chit}
\end{equation}
In this notation, the vacuum-subtracted entropy, Eq. (\ref{eq-vacsub}), is written as
\begin{equation}
\hat S_C = -\mathrm{tr}_T\, \rho_T\log \rho_T + \mathrm{tr}_T\,\chi_T\log \chi_T~.
\end{equation}
The subscript $T$ (or $\not T$) on the trace indicates that the trace is taken over the Hilbert space factor associated with the observation interval (or its complement).

\begin{figure}[t]
	\centering
	\includegraphics[width=\columnwidth]{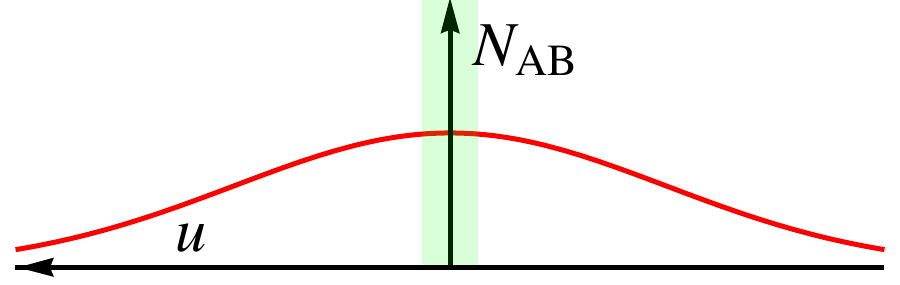}
	\caption{A short observation (green shaded rectangle) cannot distinguished the reduced graviton state from the vacuum reduced to the same region. The graviton delivers no information to this observer.}\label{fig:short}
\end{figure}

\paragraph*{Short Observation Regime}
We begin by considering the case where $\lambda\gg T$. In this regime, the observer has access to a region occupied by the graviton wavepacket, but much smaller than the wavepacket (Fig.~\ref{fig:short}). The Boundary QBB implies
\begin{equation}
\hat S_C \lesssim O(\hat {\cal T} T^2/\hbar) \sim O(T^2/\lambda^2)~,
\label{eq-scsmall} 
\end{equation}
so the upper bound vanishes quadratically with $T/\lambda$. 

To understand this result, it is instructive to return to the bulk and consider the case of a scalar field wavepacket passing through $H(u_p)$. In this setting, the entropy can be computed explicitly; and the bound has been proven~\cite{BouCas14a,BouCas14b}. A beautiful explanation of the vanishing of the information content was given by Casini~\cite{Cas08}, building on pioneering work of Marolf, Minic, and Ross~\cite{MarMin04}. 

To an observer with access to a finite or semi-infinite region, the vacuum (restricted to this region) is a noisy state. For example, in the simplest case of a semi-infinite region (Rindler space), the restricted vacuum is a thermal state. Further restrictions only make the fluctuations larger. This means that the global vacuum restricted to the interval $(-T,T)$ is a state in which thermal-like excitations with energy up to order $\hbar/T$ are unsuppressed. This energy is larger, by a huge factor $\lambda^2/T^2$, than the total energy of the graviton in this region. This is the physical origin of Eq.~(\ref{eq-scsmall}): because of thermal noise, states with and without the graviton wavepacket cannot be distinguished by an observer with access to a small subregion of the wavepacket. In short, the vacuum-subtracted entropy is a physical quantity that correctly captures how much information can be gained by a given observer.

We can also shift the observation interval so that it fails to overlap with the graviton. This is analogous to a case of classical Bondi news studied in Sec.~\ref{sec-state}, and it gives the same result: In this case it does not matter how long or short the observation is; if it does not overlap with the news, then upper bound vanishes.
\begin{figure}[h]
	\centering
	\includegraphics[width=0.7 \columnwidth]{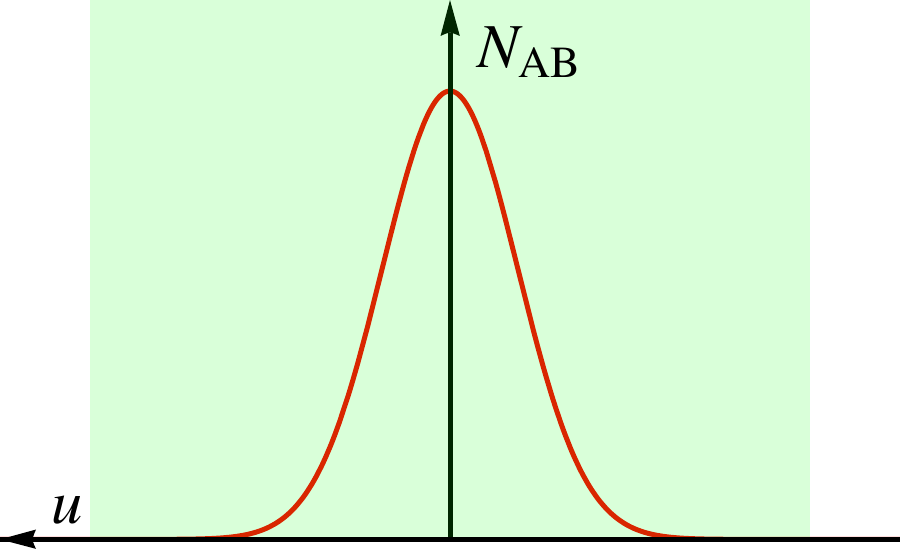}
	\caption{A long observation (green shaded rectangle) can distinguish the reduced graviton state from the reduced vacuum. The graviton carries information to this observer.}\label{fig:long}
\end{figure}

\paragraph*{Long Observation Regime}
Next, let us consider the case where the observer has access to a region that includes the whole wavepacket:  $T\gg \lambda$ (Fig.~\ref{fig:long}). In the long-observation regime, the experiment begins well before the graviton starts arriving, and ends well after. From Eq.~(\ref{eq-calt}) we see that the energy density $\hat{\cal T}$ scales as $\lambda^{-2}$. The Boundary QBB, Eq.~(\ref{eq-qbbscri0}), evaluates to 
\begin{equation}
\hat S_C\lesssim O(\hat{\cal T} T\lambda/\hbar) \sim O\left(\frac{T}{\lambda}\right)~,
\label{eq-sctl}
\end{equation}
as the integral has support only only on the central interval of size $O(\lambda)$ where $\hat g\sim O(T)$. Since we have $T\gg\lambda$, Eq.~(\ref{eq-sctl}) is consistent with the ability of the observer to extract information from the graviton.

We may specify a ``soft limit'' of the long-observation regime as follows: Let $T=\alpha \lambda$, with $\alpha\gg 1$ fixed. Then we take $\lambda$ to become as large as we like, while the experiment always lasts longer than the wavepacket. We note that the upper bound remains fixed in this limit, at $O(\alpha)\gg 1$. We can tighten the upper bound to $O(1)$ while remaining marginally within the long-observation regime by taking $\alpha\sim 1$. 

We can gain further intuition by returning to the bulk and considering the same graviton as it crosses a planar light-sheet $H(u_p)$. It induces focussing as $d\theta/dw\sim O(G\hbar/(A\lambda^2))$, were $A$ is the transverse area on which the wavepacket has support. Integrating twice along the light-rays and once transversally, we see that the area loss between the two ends of the wave packet is of order the Planck area for $\alpha\sim 1$. Thus, a single quantum induces loss of about a Planck area in planar light-sheets, independently of wavelength~\cite{Bou03}. Hence the bound on its entropy is of order unity. 

We will not try to compute the entropy of the graviton directly, but we expect it to be of order unity. To see this, let us again consider instead a scalar field wavepacket, for which the QBB has been proven~\cite{BouCas14a,BouCas14b}. We understand the presence of nonzero entropy: the experiment can access the whole wavepacket, and the excitation can be distinguished reasonably well from the thermal noise that pollutes any finite-duration measurements~\cite{MarMin04,Cas08}. Thus as $\alpha\sim 1$, the bound becomes approximately saturated at the order-of-magnitude level. 

\subsubsection{Boundary Generalized Second Law}

Finally, let us consider an observer with access to a semi-infinite region above some cut $\hat\sigma_2$ of ${\cal I}^+$. The bound that applies to this case is the integrated Boundary GSL, Eq.~(\ref{eq-gslintscric}):
\begin{equation}
\hat S_C[\hat\sigma_2]\leq \Delta K[\hat\sigma_2]~,
\label{eq-gslbig2}
\end{equation}
where
\begin{equation}
\Delta K[\hat\sigma_2]\equiv \frac{2\pi}{\hbar}\!\int_{\hat\sigma_2}^{\infty}\!\!\! d^2\Omega\, du\, 
[u-u_2(\Omega)]\, \hat{\cal T}
\label{eq-modham}
\end{equation}
is the modular Hamiltonian.

We stressed earlier that all real experiments are finite. Nevertheless, the above bound is a useful approximation for long but finite observations: first specify the global state, which must obey fall-off conditions~\cite{ChrKla93} on the news. Then restrict to an interval $(u_2,T)$ such that $T$ lies far inside the future region with essentially no news, and consider the QBB for this interval. Since the slope of $\hat g$ is unity near the lower end of the interval, and since $\hat S_C$ will no longer depend on $T$ in this regime, the Boundary QBB reduces to Eq.~(\ref{eq-gslbig2}).

\begin{figure}[h]
	\centering
	\includegraphics[width=0.7\columnwidth]{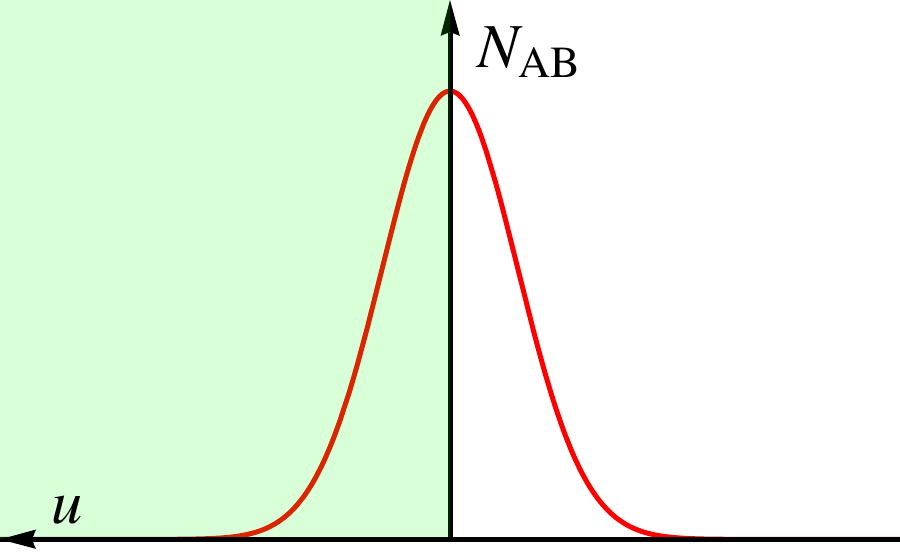}
	\caption{A graviton conveys $O(1)$ information as long as it has appreciable support in the region of observation.}\label{fig:semi}
\end{figure}

Let us apply Eq.~(\ref{eq-gslbig2}) to a graviton wavepacket with support in the region $(-\lambda,\lambda)$. First suppose that the cut $\hat\sigma_2$ lies, say, around the center of the wavepacket, as depicted in Fig.~\ref{fig:semi}. By the previous paragraph, the results will be the same as for the QBB in the regime $\alpha\sim 1$: the asymptotic geometry can be distinguished from Minkowski space, and the upper bound will be of order unity. On the other hand, if we shift the wavepacket so as to lie entirely prior to $\hat\sigma_2$, then the upper bound vanishes. 

We can also consider the differential version of the Boundary GSL, which can be written as
\begin{equation}
-\frac{1}{\delta\Omega} \frac{d}{du}\hat S_C[\hat\sigma_2;\Omega] \leq 
\frac{2\pi}{\hbar}\int_{\hat\sigma_2}^\infty du~ \hat{\cal T}~.
\label{eq-gslscri02}
\end{equation}
This vanishes if the news has no support in the region above the cut $\hat\sigma_2$. Thus, for the case of news that arrives entirely prior to $\hat\sigma$, the upper bounds on the entropy, and on its variation under deformations of $\hat\sigma$, both vanish. This is the same behavior we encountered for the classical case in Sec.~\ref{sec-state}.

In the case where a graviton wavepacket lies partially or entirely above the cut (Fig.~\ref{fig:semi}), we see that the derivative of $\hat S_C$ is bounded by the energy of the wavepacket. This is nonzero for any finite $\lambda$. We note that the upper bound depends on the energy, not on the gravitational memory created by the wavepacket. Therefore, the upper bound on the derivative of $\hat S_C$ vanishes in the soft limit as $\lambda$ becomes large, even though $\Delta C_{AB}$ remains fixed in this limit. Thus, the differential Boundary GSL implies that the variation in entropy of the region above a cut $\hat\sigma$, under fixed length deformations of $\hat\sigma$, is insensitive to the addition of gravitons of much greater wavelength.

\acknowledgments
We thank C.~Akers, H.~Casini, Z.~Fisher, E.~Flanagan, S.~Leichenauer, A.~Levine, A.~Moghaddam, R.~Wald, A.~Wall, and especially A.~Strominger for discussions and correspondence. This work was supported in part by the Berkeley Center for Theoretical Physics, by the National Science Foundation (award numbers 1214644, 1521446, and 1316783), by FQXi, and by the US Department of Energy under contract DE-AC02-05CH11231.

\bibliographystyle{utcaps}
\bibliography{all}
\end{document}